\documentclass[
 reprint,showkeys,showpacs,preprintnumbers,
 amsmath,amssymb,
 aps,
 prl,
 longbibliography,
 lengthcheck,%
]{revtex4-1}

\usepackage{graphicx}
\usepackage{bm}
\usepackage{amsmath}
\usepackage[bookmarks=false]{hyperref}
\usepackage{color}
\usepackage{dcolumn}

\usepackage[normalem]{ulem}

\begin{document}

\title{
Reference frame independent quantum key distribution server
\\with telecom tether for on-chip client
}
\author{P.~Zhang$^{1,2}$}
\author{K.~Aungskunsiri$^{1}$}
\author{E.~Mart\'{i}n-L\'{o}pez$^{1}$}
\author{J.~Wabnig$^{3}$}
\author{M.~Lobino$^{1,4}$}
\author{R.~W.~Nock$^{5} $}
\author{J.~Munns$^{1,6}$}
\author{D.~Bonneau$^{1}$}
\author{P.~Jiang$^{1} $}
\author{H.~W. Li$^{3}$}
\author{A.~Laing$^{1}$}
\email{anthony.laing@bristol.ac.uk}
\author{J.~G.~Rarity$^{5} $}
\author{A.~O.~Niskanen$^{3}$}
\author{M.~G.~Thompson$^{1} $}
\author{J.~L. O'Brien$^{1} $ }
\affiliation{
$^{1}$Centre for Quantum Photonics, H. H. Wills Physics Laboratory \& Department of Electrical and Electronic Engineering,
University of Bristol, BS8 1UB, UK,\\
$^{2}$Department of Applied Physics, Xi'an Jiaotong University, 710049, China,\\
$^{3}$Nokia Research Centre, Broers Building, 21 J.J. Thomson Avenue, Cambridge, CB3 0FA, UK,\\
$^{4}$Centre for Quantum Dynamics \& Queensland Micro and Nanotechnology Centre, Griffith University, Brisbane, QLD 4111, Australia,\\
$^{5}$Department of Electrical and Electronic Engineering, University of Bristol, BS8 1UB, UK\\
$^{6}$Bristol Centre for Functional Nanomaterials, Centre for NSQI, University of Bristol, BS8 1FD, UK
}

\begin{abstract}
We demonstrate a client-server quantum key distribution (QKD) scheme, with large resources such as laser and detectors situated at the server-side, which is accessible via telecom-fibre, to a client requiring only an on-chip polarisation rotator, that may be integrated into a handheld device.  The detrimental effects of unstable fibre birefringence are overcome by employing the reference frame independent QKD protocol for polarisation qubits in polarisation maintaining fibre, where standard QKD protocols fail, as we show for comparison.  This opens the way for quantum enhanced secure communications between companies and members of the general public equipped with handheld mobile devices, via telecom-fibre tethering.
\end{abstract}

\maketitle
The principle of quantum mechanics that requires microscopic systems to be changed upon observation has perplexed physicists since its formulation, yet understanding that the effect could be harnessed as a resource gave birth to the field of quantum cryptography
\cite{BB84,Ekert01,six-state1,six-state2,Lo99,Shor00,Gisin,e2s1,c2c1,c2c2,e2s2,e2s3,c2c3}.
Quantum enhanced security in communication is available through quantum key distribution (QKD), which exploits the behaviour of single photons to allow two parties to exchange the binary string, or key, that is used in the encryption of sensitive information.  Now implementable with current technologies \cite{Scarani09,PRAC1,PRAC2}, QKD has matured to the stage where it is moving from research laboratories towards commercial applications \cite{gi-np-1-165}.  Here we demonstrate the feasibility of equipping mobile communication devices with quantum cryptographic capabilities by making a QKD server accessible over a telecom-fibre, to which a client may tether.

\begin{figure}[t]
\centering
\includegraphics[trim=0 0 0 0, clip,width=.91\columnwidth]{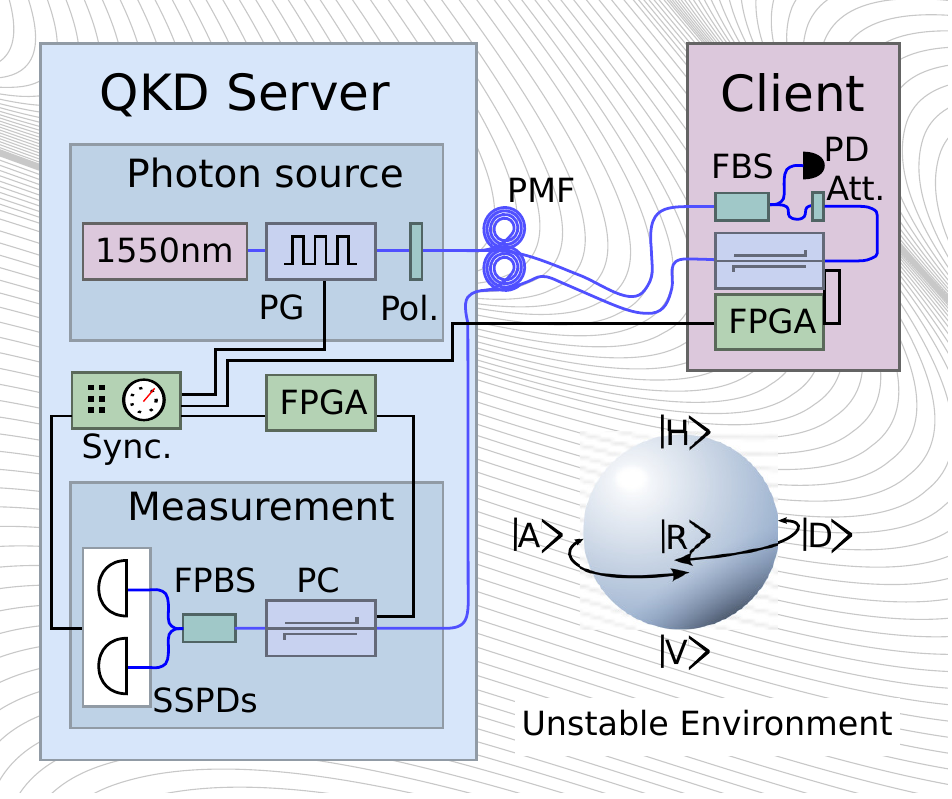}
\vspace{-3mm}
\caption{
Experimental set-up for client-server rfiQKD.  The server side holds a telecom wavelength ($1550$ nm) laser with a 1 MHz pulse generator (PG) and fixed polariser, to send light pulses to the client through a polarisation maintaining fibre (PMF).  At the client side, an integrated polarisation controller (PC) encodes qubits into the polarisation of the attenuated (Att.) light.  A fibre beam splitter (FBS) and photodetector (PD) continuously monitor power for malicious attacks.  Qubits received back at the server side are measured with a similar PC, fibre polarising beamsplitter (FPBS), and superconducting single photon detectors (SSPDs), all controlled by an electronic board synchronisation (Sync.), function programmable gate array (FPGA), and processor.
  The Bloch sphere illustrates the effects of an unstable environment on polarisation.}
\label{fgSchematic}
\vspace{-7mm}
\end{figure}

The transmission of quantum information, suitable for QKD, normally requires that a state sent by \emph{Alice} is faithfully received by \emph{Bob}, unless an eavesdropper, \emph{Eve}, observes the state and reveals her presence as an otherwise unexplainable disturbance.  Yet even in the absence of Eve, an unstable fibre communication link or instability in the sending and receiving apparatus, is equivalent to an unknown or varying reference frame, and has the effect of unhelpfully transforming the states that Bob receives \cite{arXiv-1306-6134}.

Attempts to overcome potential reference frame misalignment by encoding qubits within larger systems \cite{DFS0,DFS1,DFS2} require creation, manipulation, and detection of many-photon entangled states, which is technically challenging and very loss-sensitive.  Encoding information into the modes available from the transverse spatial profile of light \cite{DFS3,DFS4,DFS5, da-natcom-3-961}, may facilitate communication between misaligned parties, through air or vacuum, but encounters problems such as mode dispersion when transmitting through fibre.  Protocols that exploit the arrival time of photons as a logical basis necessitate stable interferometers to perform encoding and decoding, requiring active stabilisation in fibre  \cite{ma-prl-93-180502}, or precise temperature regulation in on-chip QKD with highly asymmetric interferometers \cite{ta-oe-16-11-354}.

An alternative time-multiplexing scheme is the \emph{plug and play} system which sends each light pulse back and forth along the same fibre, with the aid of a Faraday mirror, to cancel the effects of birefringence \cite{mu-apl-70-793, zb-el-33-586, ri-el-34-2116, st-njp-4-41-1, er-njp-12-063027}.  Interferometric stability results form both halves of the time-split pulse retracing each others path.  A drawback to this double-pass arrangement is the potential for an increased error rate due to Rayleigh backscattering, which requires the addition of a storage line to hold a train of pulses which must complete a round trip before the next train is sent.  A further stability constraint, related to the length of transmission, is that fluctuations in the fibre and interferometer should be slow on the time-scale of the double pass.

The reference frame independent QKD protocol (rfiQKD) \cite{RFI1, sh-njp-12-123019, RFI2, URFI}, deployed here, operates between unknown and changing reference frames, is independent of any particular choice of information encoding, requires no entanglement, and is implementable with two-level systems encoded onto single photons that may be approximated with weak coherent laser pulses, making it intrinsically practical.  The protocol will generate a secret key as long as the rate of change between reference frames is slow on the rate of particle repetition.

With freedom to choose the physical two-state encoding and no requirement for phase stability, the rfiQKD protocol allowed us to exploit commercially available lithium niobate integrated polarisation controllers \cite{bo-prl-108-053601} to implement QKD with photon-polarisation qubits over an unstablised fibre link.  Larger resources, such as the photon source and superconducting detectors \cite{ta-apl-96-221109, do-apl-93-131101}, are situated on Bob's side, which can be regarded as the \emph{server side}, while Alice, as the \emph{client}, requires only the capability to perform single qubit operations.  The scenario is one in which the client tethers a hand held device, with an integrated photonic chip, to a telecom-fibre to receive dim laser pulses from the QKD server, which the client attenuates to the single photon level, before encoding each pulse with a qubit of information for return transmission to the server, along a different fibre.

Here, we demonstrate a stable, constant, and continuously positive secret key rate over the unstablised fibre link using the rfiQKD protocol, while the rate for the BB84 QKD protocol falls.  We go on to show that our rfiQKD system automatically and passively recovers from the deliberate introduction of large amounts of noise in the form of rapid fluctuations.

Although formulated as an entanglement-driven protocol, rfiQKD can be implemented with weak coherent states that sufficiently approximate single photons, where Alice randomly prepares and sends the polarisation states $\{D/A, R/L, H/V\}$ corresponding to the eigenvectors of the Pauli matrices, which we label as $\{X,Y,Z\}$ \cite{foot1}.  In our experiment, the requirement of one known and stable basis, used to encode the key, is fulfilled with the horizontal ($H$) and vertical ($V$) polarisation states, that are preserved throughout transmission in polarisation maintaining fibre (PMF).  The other four states, superpositions of $H$ and $V$ used to guarantee security, are unhelpfully transformed by phase fluctuations in PMF due to environmental influences on the birefringence of the fibre; while this effect is troublesome for other protocols, rfiQKD operates in the presence of phase drifts that are slow on the repetition rate of sent photons.  For fluctuations sufficiently rapid to force protocol failure, rfiQKD will automatically recover in calmer periods without the need for re-alignment, as we demonstrate.

The operation of the protocol, with its phase invariant security measure, works as follows: Expressing Alice and Bob's measurement bases as $Z_A=Z_B$, $X_B=\cos(\beta) X_A + \sin(\beta) Y_A$ and $Y_B=\cos(\beta) Y_A - \sin(\beta) X_A$, where $\beta$ slowly changes with time in an unknown way, Alice randomly prepares quantum states which she sends to Bob, who measures in his randomly chosen basis; later they publicly reveal their choice of bases.  The raw key is obtained when they both measure in the $Z$ direction, providing a quantum bit-error rate
\begin{equation}
Q=\frac{1-\langle Z_A Z_B\rangle }{2}.
\label{eqQ}
\end{equation}

The other two slowly rotating bases are used to estimate the knowledge of a potential Eve. The quantity
\begin{eqnarray}
&&C=\langle X_A X_B\rangle ^2 + \langle X_A Y_B\rangle ^2 \notag \\ 
&&\ \ \ \ \ \ \ \ + \langle Y_A X_B\rangle ^2 + \langle Y_A Y_B\rangle ^2,
\label{eqC}
\end{eqnarray}
is independent of the relative angle $\beta$. If there is no Eve and the communication channel is ideal with a fixed (although unknown) phase, then the correlation function $\langle Z_A Z_B\rangle $ is equal to $1$ whereas $\langle X_A X_B\rangle$, $\langle X_A Y_B\rangle $, $\langle Y_A X_B\rangle $ and $\langle Y_A Y_B\rangle $ each take take a constant value between $-1$ and $1$ determined by $\beta$. Also, $\langle Z_A X_B\rangle $, $\langle Z_A Y_B\rangle $, $\langle X_A Z_B\rangle $ and $\langle Y_A Z_B\rangle $ should be zero as $X$, $Y$, and $Z$ are mutually unbiased. With Z bases aligned, $Q=0$ and $C=2$ will be achieved, but in a realistic implementation, $Q$ will be greater than zero and $C$ will be less than 2.

The correlation functions involved in (\ref{eqQ}) and (\ref{eqC}) are calculated from the rates of photon detections by assigning positive or negative signs to correlated or anti-correlated detections respectively.  For example, if Bob labels his pair of detectors as $b=\{0,1\}$, while Alice labels the pair of states she sends (within a particular basis) as $a=\{0,1\}$ then the $\langle A B\rangle$ expectation value is calculated from the number of detector clicks
$n_{a\,b}$ as
$(n_{00} + n_{11} - n_{01} - n_{10}) / (n_{00} + n_{11} + n_{01} + n_{10})$,
which is essentially the normalised difference of correlated and anti correlated detections.

The security proof of the rfiQKD protocol \cite{RFI1} shows that when $Q\lesssim15.9\%$, Eve's information is given by
\begin{align}
E(Q,C) = \frac{h}{2}(1+(1-Q)u_{max} \, + \, Q\,v(u_{max}))
\end{align}
where
\begin{align}
u_{max} &= \min [\frac{\sqrt{C/2}}{1-Q},1], \notag \\
v &=\frac{1}{Q}\sqrt{C/2-(1-Q)^2 u_{max} ^2}, \notag
\end{align}
and the binary entropy
\begin{align}
h(x)=-x\log _2 (x) -(1-x)\log _2 (1-x). \notag
\end{align}
The secret key rate is given by
\begin{equation}
r=1-h(Q) - E(Q,C).
\end{equation}

The experimental setup is shown in Fig.~\ref{fgSchematic}.  At the server side, light from a 1550 nm continuous-wave laser source is sent through a pulse generator (PG) to produce pulses of 100 ns width with a repetition rate of 1 MHz \cite{foot2}, which are filtered by a horizontal polariser and transmitted to the client through PMF.  At the client side, a fibre beam splitter and photodetector expose hypothetical attacks (for example, \cite{ly-np-4-686}). 
A $\approx75$ dB attenuator reduces the light intensity to the single photon level of $\approx0.1$ photons per pulse so that the probability of more than one photon per pulse $\approx0.005$.  The client randomly prepares among the six polarisation states $\{D,A,R,L,H,V\}$ using a LiNbO$_3$ \cite{Thaniyavarn85APL,Thaniyavarn85OL} polarisation controller (PC) commanded from a field programmable gate array (FPGA) and associated driver circuits.  Photonic qubits are returned to the server, along another unstabilised 5 m length of PMF, where a similar PC and FPGA, together with a fibre polarisation beam splitter and superconducting single photon detectors, perform projective measurements chosen randomly among the three relevant bases.

\begin{figure}[t]
\centering
\includegraphics[trim=5 0 0 0, clip,width=1\columnwidth]{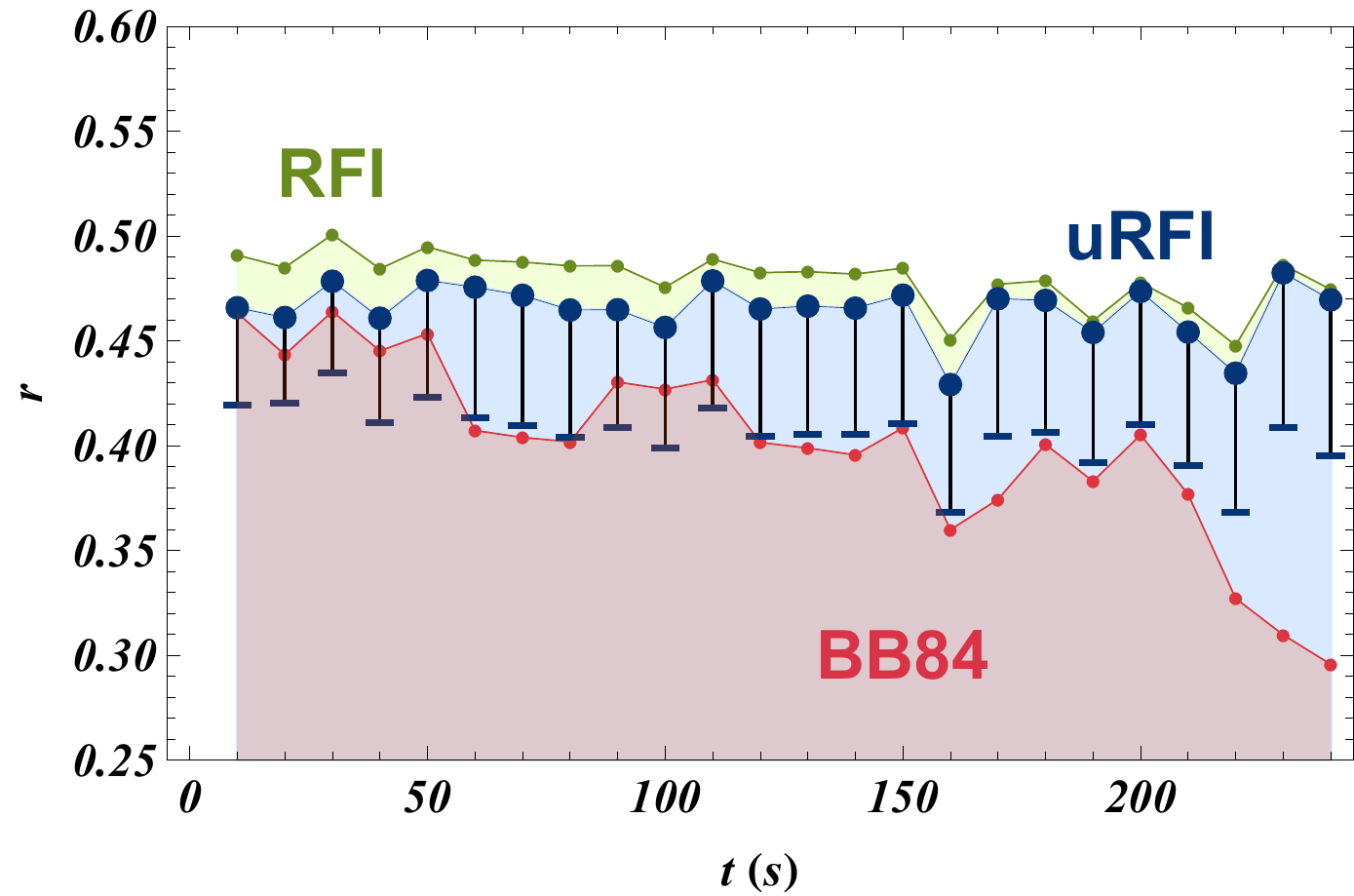}
\caption{
Experimental data for secret key rate fraction $r$ showing robustness to drift.
Data is initially collected in the situation of well aligned client-server reference frames, but the unfixed PMF quantum channel is subject to ambient environmental influences, which effect a slowly varying reference frame.
While the key rate for the rfiQKD protocols is constant, for BB84 it suffers a fall as the alignment drifts.  Lower bounds on the secret key rate from the urfiQKD analysis are shown.
}
\label{fgQuiet}
\vspace{-2mm}
\end{figure}

All system elements are synchronised to the SYNC-FPGA platform, which allows for precise timing of all stages within the transmission period.  Each repetition begins with an optical pulse from the server side PG; when the pulse arrives at the client PC, it is set to prepare a particular polarisation state; then, timed appropriately for the return of the pulse, the server PC is set to measure in a particular basis; finally, the state of the detectors is recorded by the FPGA.

In addition to demonstrating the feasibility of QKD between telecom-fibre linked integrated photonic devices in the described client-server scheme, we aim to show two features of the rfiQKD protocol that are particularly relevant: robustness to phase drift, inevitable in long range fibre; and automatic passive recovery from rapid noise.

We also present analysis for the uncalibrated variant \emph{urfiQKD} protocol which assumes not only unaligned reference frames, but also removes the assumption for alignment within a reference frame, allowing for non-orthogonality within a basis and mutual \emph{bias} between bases, and differing detector efficiencies that arise in a real world implementation \cite{URFI}.  This is achieved by using an explicit device model and minimising the key rate over possible model parameters.

\begin{figure}[t]
\centering
\includegraphics[trim=0 0 0 0,width=\columnwidth]{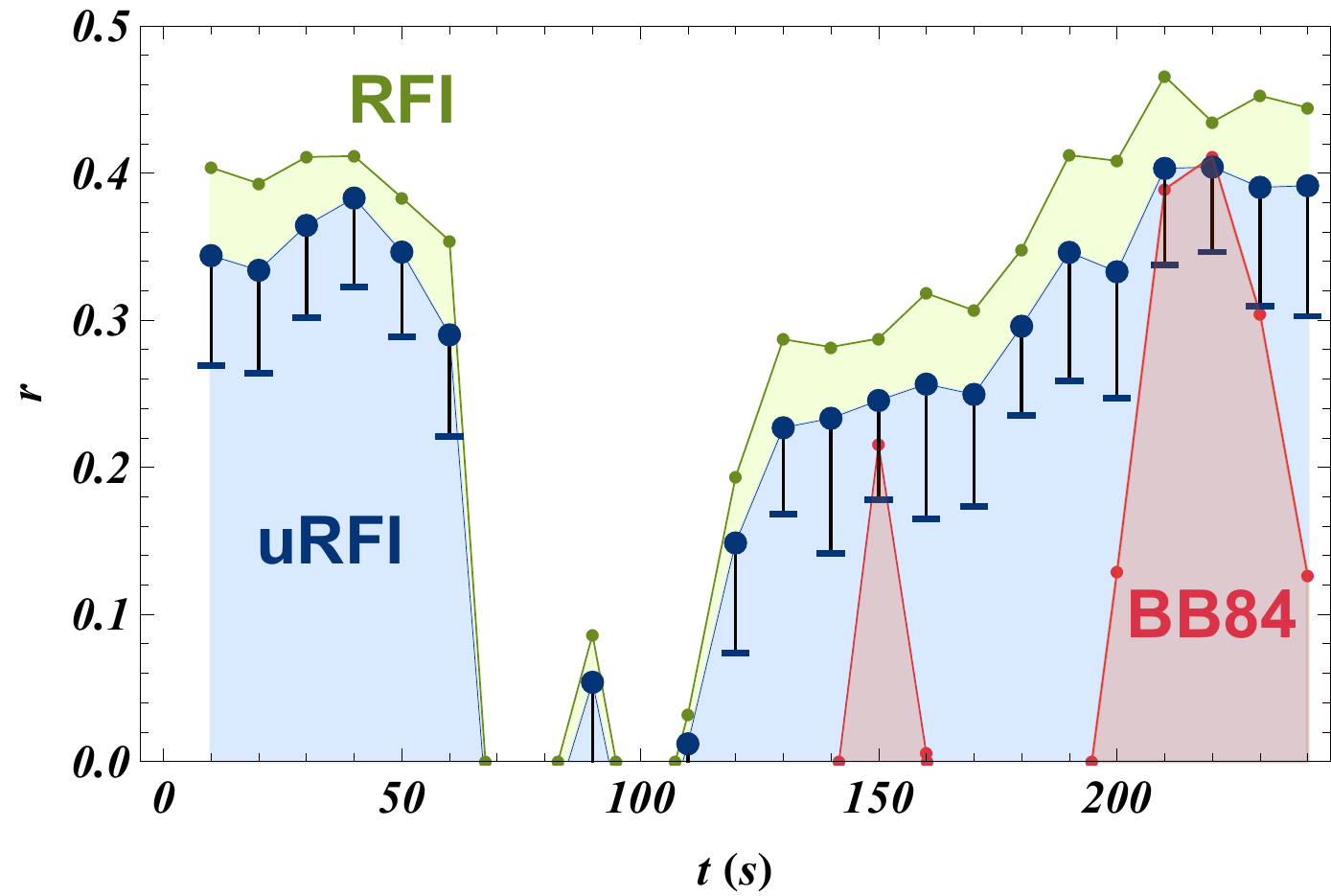}
\caption{
Experimental data for secret key rate fraction $r$ showing automatic recovery from rapid noise.
During the initial $60$s, the unaligned and slowly varying reference frames result in BB84 failure while the rfiQKD protocols operate.
Between $t = 60$s and $t=120$s we deliberately introduced rapid PMF deformations force an rfiQKD failure.
However, an automatic and passive revival of $r$ for the rfiQKD protocols is observed during the subsequent calm period.
}
\label{fgNoisy}
\vspace{-2mm}
\end{figure}

We demonstrated \emph{drift robustness} of the rfiQKD and urfiQKD generated key rates in comparison with that of BB84, beginning key exchange with well aligned client-server reference frames, so that states prepared by the client PC had a high fidelity with projectors determined by the server PC.  The PMF quantum channel was unfixed but undisturbed for the duration of the key exchange, subject only to ambient environmental influences.  Figure.~\ref{fgQuiet} shows the secret key rate fraction, $r$, as a function of time for the BB84, rfiQKD, and urfiQKD protocols.  The duration of key exchange is 240 s, so each of the 24 points corresponds to data integrated over 10 seconds, which is long enough to collect sufficient data to produce small error bars, corresponding to a precision of 3 standard deviations, but short enough to avoid significant depletion of the value of $r$, from integration over largely different values $\beta$.  For clarity, these error bars are not displayed, instead we show the lower bound on $r$ from the urfiQKD analysis.  It can be seen that while the secret key rate for BB84 falls as a function of time, the rfiQKD protocols do not.

We demonstrated the \emph{automatic, passive recovery} capability of our system after periods of rapid and substantial noise that force a protocol failure.  Figure.~\ref{fgNoisy} shows $r$ as a function of time for 24 points, each corresponding to 10 seconds of data, as before.  During the initial 60 s of key exchange, the PMF quantum channel is unfixed and undisturbed but unaligned so that the BB84 protocol immediately fails.  However the changes resulting from the ambient environment are slow enough for the rfiQKD protocols to operate successfully.  Between $t=60$ s and $t=120$ s we deliberately introduced a large amount of noise by continually and significantly deforming the PMF to simulate a rapidly changing reference frame, forcing $r$ to fall below zero.  At $t=120$ s, the noise ceases and as the PMF relaxes from the mechanical strains, a positive key rate automatically returns and achieves initial values for the rfiQKD protocols.  In contrast, BB84 achieves a positive $r$ only in brief transitionary periods of near-alignment.  Again, the lower bound on $r$ for the urfiQKD analysis is displayed.

In conclusion, we demonstrated a client-server QKD protocol where all large resources reside at server side, and the client requires only an integrated photonic device that could be further integrated into a hand held communication device. The key is exchanged though a PMF telecom-fibre tether using the rfiQKD protocol, that is shown to be passively robust to typical environmental drift effects and can automatically recover from large noise levels to re-establish QKD in calmer periods with no requirement for alignment. The results significantly broaden the operating potential for QKD outside of the laboratory and pave the way for quantum enhanced security for the general public with handheld mobile devices.

Future directions are to adapt the system for time multiplexing with the rfiQKD protocol which avoids the requirement for temperature-stabilised on-chip asymmetric Mach Zehnder interferometers.  Such a system would be well suited to take advantage of existing (non PMF) telecom-fibre networks, without the need for double passing and the inheritance of associated problems.  The challenge is to then increase the distance over which secret key may be exchanged in telecom-fibre.

\section{Acknowledgements}
This work was supported by EPSRC, ERC, QUANTIP, PHORBITEC and NSQI. P. Z. acknowledges support from the Fundamental Research Funds for the Central Universities, the Special Prophase Project on the National Basic Research Program of China (Grant No.2011CB311807), and the National Natural Science Foundation of China (Grant No. 11004158).  J.M. acknowledges EPSRC grant code EP/G036780/1.  J.L.OB. acknowledges a Royal Society Wolfson Merit Award and a Royal Academy of Engineering Chair in Emerging Technologies.


\begin{thebibliography}{99}
\bibitem{foot1} As is typical, $\{D,A,R,L,H,V\} \equiv $ \{diagonal, anti-diagonal, right circular, left circular, horizontal, vertical\}.
\bibitem{foot2} The initial $1$MHz pulse rate is attenuated at client-side to an average intensity of $\approx 0.1$ photon/pulse; this and the efficiency of the SSPDs, at $10\%$, give a final average repetition rate of $10$kHz.
\end{thebibliography}
\end{document}